\documentstyle[aps,prl,twocolumn,floats]{revtex}
\def\pht{W}

\begin{document}

\twocolumn[\hsize\textwidth\columnwidth\hsize\csname @twocolumnfalse\endcsname

\title{ Quantum Transition between an Antiferromagnetic 
Mott Insulator and $d_{x^2 - y^2}$ Superconductor in Two Dimensions}

\author{ F.F. Assaad$^{1}$, M. Imada$^{2}$  and D.J. Scalapino$^{1}$ \\
	 $^{1}$ Department of Physics, University of California \\
           Santa Barbara, CA 93106-9530 \\
         $^{2}$ Institute for Solid State Physics, University of Tokyo,  \\
         7-22-1 Roppongi,
         Minato-ku, Tokyo 106, Japan.  }

\maketitle

\begin{abstract}
We consider a Hubbard model on a square lattice with an additional
interaction, $\pht$, which depends upon the square of a near-neighbor 
hopping. 
At half-filling and a constant value of the Hubbard
repulsion, increasing the strength of the interaction $\pht$ drives the 
system from an antiferromagnetic Mott insulator to a 
$d_{x^2 -y^2}$ superconductor. This
conclusion is reached on the basis of zero temperature quantum Monte Carlo 
simulations on lattice sizes up to $16 \times 16$.  
\end{abstract}

]

At half-filling, the two-dimensional (2D) Hubbard model  on a square 
lattice with a near-neighbor one-electron hopping t and an onsite Coulomb
interaction $U$ has an insulating ground state with long range 
antiferromagnetic (AF) order. When carriers are doped into this insulating
state, the possibility of a transition to a $d_{x^2 - y^2}$ superconducting 
state  has been extensively pursued.   
However, at energy scales and lattice sizes accessible to 
numerical simulations of the 2D Hubbard model,  no definite sign of
superconductivity has been found in the vicinity of the Mott transition
\cite{Scalapino}. An alternate approach for studying the transition 
from the Mott insulator to a 
$d_{x^2 - y^2}$ superconducting state is to keep the system at half-filling,
but add an interaction which can drive the system into a $d_{x^2 - y^2}$ 
superconducting ground state. Here, we introduce a simple form for
such an interaction which
depends on the square of a near-neighbor hopping. We note that it can formally
be obtained from a Su-Schrieffer-Heeger \cite{Su} 
interaction in the antiadiabatic limit.  
It may also be viewed as derived from purely electronic origins near 
the AF Mott insulator. 
However, our purpose is not aimed at describing the pairing mechanism
but rather at studying the transition between an antiferromagnetic Mott 
insulating state and a $d_{x^2 - y^2}$ superconducting state. In this sense,
we view the interaction as a formal way of obtaining a model which exhibits
this transition and is suitable for Monte-Carlo simulations in that there are
no fermion sign problem. 

The basic half-filled Hubbard model that we will study has the Hamiltonian
\begin{equation}
      H_U  =  -\frac{t}{2} \sum_{\vec{i}} K_{\vec{i}} + U \sum_{\vec{i}}
         (n_{\vec{i},\uparrow}-\frac{1}{2})
         (n_{\vec{i},\downarrow} -\frac{1}{2}) 
\end{equation}
with the hopping kinetic energy
\begin{equation}
 	K_{\vec{i}} = \sum_{\sigma, \vec{\delta}}
   \left(c_{\vec{i},\sigma}^{\dagger} c_{\vec{i} + \vec{\delta},\sigma} +
        c_{\vec{i} + \vec{\delta},\sigma}^{\dagger} c_{\vec{i},\sigma} \right). 
\end{equation}
Here, 
$c_{\vec{i},\sigma}^{\dagger}$ ($c_{\vec{i},\sigma}$) creates (annihilates) an
electron with {\it z}-component of spin $\sigma$ on site
$\vec{i}$, $n_{\vec{i},\sigma } =  c_{\vec{i},\sigma}^{\dagger}
c_{\vec{i}\sigma}$, and $\vec{\delta} = \pm \vec{a}_x, \pm \vec{a}_y $ where
$\vec{a}_x$, $\vec{a}_y$  are the lattice constants. The energy 
will be measured in units of $t$. 
The interaction that we will add has the form:
\begin{equation}
H_\pht  =  -\pht \sum_{\vec{i}} K_{\vec{i}}^{2}
\end{equation}
with positive $\pht$. We note that the interaction $H_\pht$  can be generated
from a  Su-Schrieffer-Heeger \cite{Su} term with Einstein oscillators:
\begin{eqnarray}
	\sum_{\langle\vec{i},\vec{j} \rangle,\sigma} 
    \vec{\lambda} \cdot \left( \vec{Q}_{\vec{i}} - \vec{Q}_{\vec{j}} \right)
  & & \left(c_{\vec{i}, \sigma}^{\dagger} c_{\vec{j}, \sigma} + {\rm h.c.} \right) +
\nonumber \\
  & & \sum_{\vec{i}} \left( \frac{\vec{P}_{\vec{i}}^2}{2M} +
        \vec{Q_{i}}^{\dagger} \frac{D}{2} \vec{Q_{i}} \right).
\end{eqnarray}
Integrating out the phonons and taking the antiadiabatic limit ($M
\rightarrow 0 $), generates $H_\pht$ with 
$ \pht = \vec{\lambda}^\dagger D^{-1} \vec{\lambda} / 2 $. Pairing mechanism
along those lines were considered in \cite{Hirsch1,Imada1}.
This derivation should be compared to the case of the attractive Hubbard model 
which may be obtained in a similar way provided that electron-phonon
interaction is chosen to be of the Holstein form \cite{Hirsch}.
In fact, one of the reasons for choosing this form is that it has a 
simple Hubbard-Stratonovitch representation which is  useful in 
constructing the Monte-Carlo simulation. 

The Hamiltonian 
\begin{equation}
\label{PHM}
        H = H_U + H_\pht
\end{equation}
has the possibility of exhibiting a
quantum transition between an AF Mott insulating state and a
superconducting phase.  When $ \pht = 0$, the half filled Hubbard model with
a finite $U$ is known to be a Mott insulator with long-range  AF  order. 
The interaction $H_\pht$ can be decomposed into 
single-particle next-nearest neighbor hopping terms, 
singlet pair-hopping  terms involving on-site singlet pairs,
triplet pair-hopping processes,  and 
terms of the form:
\begin{equation}
\label{PH}
  -\pht \sum_{\vec{i}, \vec{\delta}, \vec{\delta'}}
\left(\Delta^{\dagger}_{\vec{i}, \vec{\delta'}} \Delta_{\vec{i}, \vec{\delta}}+
 {\rm h.c.}  \right) 
\end{equation}
where $ \Delta^{\dagger}_{\vec{i}, \delta} = \left(
c_{\vec{i},             \uparrow}^{\dagger}
c_{\vec{i}+\vec{\delta},\downarrow}^{\dagger} -
c_{\vec{i},             \downarrow}^{\dagger}
c_{\vec{i}+\vec{\delta} ,\uparrow}^{\dagger} \right)/\sqrt{2} $. 
In the presence of the Hubbard  Hamiltonian $H_U$, 
the latter term (\ref{PH}) dominates the low energy physics, and 
as we will see, leads to a $d_{x^2 - y^2}$ superconducting state. 

We have carried out our simulations with a zero temperature quantum
Monte-Carlo (QMC) algorithm \cite{Koonin,Sandro}. 
At half-filling the Hamiltonian,  Eq.(\ref{PHM}),  is particle-hole
symmetric  at any $U$ and $\pht$ so that there is no sign problem and 
ground state simulations on
{\it large} lattices (up to $16 \times 16$) were carried out without
any complications. 
This statement is valid even for twisted boundary 
conditions for which
\begin{equation}
\label{Bound}
 c_{\vec{i} + L \vec{a}_x, \sigma } = \exp \left(2 \pi i \Phi/\Phi_0
\right) c_{\vec{i}, \sigma} $ and $c_{\vec{i} + L \vec{a}_y, \sigma } 
= c_{\vec{i}, \sigma},
\end{equation}
with $\Phi_0 = h c / e$ the flux quanta and  $L$ the  linear length 
of the square lattice. 
The boundary conditions given by Eq. (\ref{Bound}) 
account for a magnetic flux threading a torus on which  the lattice is 
wrapped.
To take advantage of the efficiency of a single spin-flip algorithm,
we have carried out a discrete Hubbard Stratonovitch
transformation of the $H_p$ term. The transformation produces an
overall systematic error of order $(\Delta \tau \pht)^3$  where $\Delta \tau $ 
corresponds to the imaginary time step entering  the path integral 
formulation. 
This systematic error is negligible compared to the one 
produced by the Trotter decomposition which is of 
order $(\Delta \tau )^2$. Unless mentioned otherwise, we have carried
out our simulations at $\Delta \tau t  = 0.0625$. 
The details of the algorithm will be presented elsewhere \cite{Assaad2}.

We first concentrate on the charge degrees freedom with 
$U/t=4$, and study the ground state as a function of $\pht/t$. 
In order to do so, we have computed the 
ground state energy as a function of the twist in the boundary
condition: $E_0(\Phi)$. For an insulator, the wave function is localized
and hence, an exponential decay of $ \Delta E_0(\Phi) \equiv E_0(\Phi) -
E_0(\Phi_0/2)$ as a function of lattice size is expected \cite{Kohn}. 
In the spin density wave (SDW)
approximation for the half-filled Hubbard model, one obtains 
$ \Delta E_0(\Phi)  = \alpha(\Phi) L \exp \left( -L/\xi \right)$ where
$\xi$  is the localization lenght of the wavefunction. 
On the other hand, for a superconductor, $ \Delta E_0(\Phi) $ 
shows anomalous flux quantization: $ \Delta E_0(\Phi)$ is a
periodic function of $\Phi$ with period $ \Phi_{0}/2 $ and a 
non vanishing energy barrier is to be found between the flux minima 
\cite{Byers,Yang,Assaad1} so that $\Delta E_0(\Phi_0/4)$ remains finite as
$L \rightarrow \infty $.  
Fig. 1a shows $\Delta E_0(\Phi_0/4)$  versus $1/L$ for various values 
of $\pht/t$. 
One observes a change in the size-scaling of $\Delta E_0( \Phi_0/4)$  
as $\pht/t$ decreases from $\pht/t = 0.5$ to $\pht/t = 0.22$. From these
measurments, we estimate that the change occurs in the vicinity of 
$\pht/t  = 0.3$.
For values of $\pht/t <   0.3$ $\Delta E_0(\Phi_0/4)$ is consistent 
with the SDW form  whereas for $\pht/t \geq  0.33$ $\Delta
E_0(\Phi_0/4)$ may be fitted to a $1/L$ form and scales to a finite
value. The extrapolated value of $\Delta E_0(\Phi_0/4)$ versus $\pht/t$ 
is plotted in Fig. 1b  and the 
quantum transition between a Mott insulator and superconductor occurs at
$\pht/t \sim 0.3 $.

In order to determine the symmetry of the order parameter in the
superconducting state, we have calculated pair-field correlations 
in the $s$ and $d_{x^2 - y^2}$ channels: 
\begin{equation}
P_{d,s} (\vec{r}) = \langle \Delta_{d,s}^{\dagger}(\vec{r})
\Delta_{d,s}(\vec{0}) \rangle 
\end{equation}
with 
\begin{equation} 
\Delta_{d,s}^{\dagger}(\vec{r})  = 
\sum_{\sigma,\vec{\delta}}  f_{d,s}(\vec{\delta}) 
\sigma c^{\dagger}_{\vec{r},\sigma}
c^{\dagger}_{\vec{r} + \vec{\delta},-\sigma} . 
\end{equation}
Here, $f_{s}(\vec{\delta}) = 1$ and $f_{d}(\vec{\delta}) = 1 (-1)$ 
for $\vec{\delta} = \pm \vec{a_x}$ ($\pm \vec{a_y})$.
Fig. 2 shows plots of $P_{d,s} (L/2,L/2)$ for $\pht/t = 0.6$ where the system 
is superconducting and $\pht/t = 0.1$ where it is not. 
At $\pht/t = 0.6$, the dominant signal
at long distances ($L = 16$) is obtained in the $d_{x^2 - y^2}$ channel. 

At the mean-field level, the symmetry of the order parameter will determine
the  functional form of the single particle occupation number, $n(\vec{k})$.
For a $d_{x^2 - y^2}$ superconductor the BCS result yields: 
\begin{equation}
\label{nk}
n(\vec{k}) = 1 + \frac{\epsilon_{\vec{k}}}{
\sqrt{ \Delta_{\vec{k}}^2 +  \epsilon_{\vec{k}}^2 } }
\end{equation}
where $ \epsilon_{\vec{k}} = -2 t \left(\cos(k_x) + \cos(k_y) \right)$
and $ \Delta_{\vec{k}} = 
\Delta_0 \left(\cos(k_x) - \cos(k_y) \right)$. 
As apparent from Eq. (\ref{nk}) in the $ \vec{k} = k(1,1)$ direction
the $d_{x^2 - y^2}$ gap vanishes and
$ n(\vec{k}) $ shows a jump at the Fermi energy, whereas in the
$ \vec{k} = k(1,0) $ direction  $ n(\vec{k}) $ is a smooth function of
$ k$.  Precisely this behavior in $ n(\vec{k})$ may be detected in the QMC
data at $\pht/t = 0.6$ as shown in Fig. 3a. For comparison, we have plotted 
$n(\vec{k})$ at $\pht =0$ where it is expected to scale to a smooth function  in
the thermodynamic limit (see Fig.3b). 

We now consider the spin degrees of freedom. We have computed the real
space spin-spin correlations: 
\begin{equation} 
S(\vec{r}) = \frac{4}{3} \langle \vec{S}
(\vec{r} ) \vec{S} (\vec{0} ) \rangle
\end{equation}
where 
$\vec{S}\left(\vec{r} \right) $ is the spin operator at site $\vec{r}$.
For values of $\pht/t < 0.3 $ and lattice sizes ranging from 
$L=4$ to $L=12$,  $S(L/2,L/2) $,  may be fitted to a 
$1/L$ form  and scales to a finite value, as shown in Fig. 4a. 
We therefore conclude that long range
AF order is present for $\pht/t < 0.3$. 
The associated staggered moment, $ m = \lim_{L \rightarrow \infty}
\sqrt{3 S(L/2,L/2) } $,
is plotted in Fig 4b. The data is consistent with a continuous decay
of $m$ as $\pht/t$ increases towards $0.3$.  
At $\pht/t = 0.3$, we were unable to distinguish
$m$ from zero within our statistical uncertainty. Hence, we conclude
that long-range AF order vanishes at $\pht/t \sim 0.3 $. 
Therefore, within our numerical resolution, the antiferromagnetic
transition point is not separated from the superconductor-insulator
transition point.
Well within the $d_{x^2 - y^2}$ superconducting phase the
spin-spin  correlations remain sizable. In fact, at $\pht/t = 0.6$,
lattice sizes ranging from $L=4$ to $L=16$,  $S(L/2,L/2 ) $ scales
as $L^{-\alpha}$ with $\alpha = 1.16 \pm 0.01 $ as shown in the
inset of Fig. 4a.  This slow decay of the spin-spin correlations in the 
superconducting state arise because of the nodes in the $d_{x^2 - y^2}$ 
gap. 
Following Ref. \cite{Nejat}, one can approximate the spin
susceptibility, $\chi(\vec{q}, i\omega_m)$, in the superconducting state
by inserting the irreducible
BCS spin susceptibility, $\chi_0(\vec{q}, i\omega_m)$, in the random phase 
approximation (RPA) form of  $\chi(\vec{q}, i \omega_m)$:
$\chi_{RPA}(\vec{q}, i \omega_m) = \chi_0(\vec{q}, i \omega_m)/ (1 - U
\chi_0(\vec{q}, i \omega_m) )$. Here, $\omega_m$ corresponds to
Matsubara frequencies.  Within this approximation and at half-band 
filling,  the spin-spin correlations for a $d_{x^2 - y^2}$ 
superconducting order parameter show a powerlaw decay. In contrast, 
for an $s-$wave order parameter, an exponential decay is obtained.

We have found a simple model, Eq. (\ref{PHM}), which exhibits a ground state
transition from an AF Mott insulator to a $d_{x^2 - y^2}$ superconductor.
Specifically, for $U/t = 4$ at half-filling ($\langle n \rangle = 1$)  we
have found that this transition occurs at $\pht_c \sim 0.3 t$.
As a natural consequence of these results, we expect
for values of $\pht < \pht_c$  the occurrence of a quantum transition
between an AF Mott insulator and a $d_{x^2 - y^2}$ superconductor as a
function of chemical potential, $\mu$. 
It has recently been concluded that for the 2D Hubbard
model ($\pht=0$),  the quantum transition between the AF 
Mott insulator and metallic state driven by the chemical potential,
belongs to a universality class characterized by an unusually 
large dynamical exponent, $ z=4 $ \cite{Imada,Furukawa,Assaad}.
The presence of the $H_\pht$ term may
by-pass this suppression of coherence observed at $\pht=0$ and
introduce another independent energy scale to drive the Mott transition.
At constant $U$ and in the $\pht$, $\mu$ plane, the Hamiltonian (\ref{PHM})
is expected to lead to a rich structure of quantum
transitions involving the Mott insulating state, the d-wave
superconducting state and metallic states as well as crossovers at 
finite temperatures.
Although the justification of the model from a microscopic point of
view is at present uncertain, it provides us with a model to study 
some of the salient ground state and finite temperature features of
a $d_{x^2 - y^2}$ superconductor at an energy scale 
accessible to numerical simulations. It also allows us to study
a quantum transition between an AF Mott insulator and a $d_{x^2 - y^2}$
superconductor.

The numerical simulations were  carried out on the FACOM VPP 
500/40 of the Supercomputer Center of the Institute for Solid State 
Physics, University of Tokyo. D.J.S. and F.F.A. have benefited 
from discussions with W. Hanke and N. Bulut.
F.F.A  thanks the Swiss National Science  foundation for financial support
under the grant number 8220-042824. M.I. thanks the Institute of Theoretical
Physics at the University of California, Santa-Barbara for hospitality and the
National Science Foundation for financial support under the grant number
PHY94-07194.  D.J.S.  and F.F.A. acknowledge partial support from
the National Science Foundation under the grant No. DMR95-27304.

\subsubsection*{Figure captions}
\newcounter{bean}
\begin{list}%
{Fig. \arabic{bean}}{\usecounter{bean}
                   \setlength{\rightmargin}{\leftmargin}}
                                                                
\item (a) $\Delta E_0(\Phi_0/4) \equiv E_0(\Phi_0/4) - E_0(\Phi_0/2)$ 
versus $1/L$  for several values of $\pht/t$. For $\pht/t < 0.3 $  the solid 
lines correspond to a least square fit of the data to the SDW form: 
$L \exp(-L/\xi)$. For $\pht/t  > 0.3 $ the QMC data is compatible 
with a $1/L$ scaling. The solid lines are least square fit to this form. 
(b) Extrapolated value of $\Delta E_0(\Phi_0/4)$ versus $\pht/t$.
The solid line is a guide to the eye.  

\item  $d_{x^2 - y^2}$ (triangles) and $s$-wave (circles) 
pair-field correlations  versus $1/L$.

\item (a) $n(\vec{k})$ at $\pht/t = 0.6$, $U/t=4$ and $\langle n \rangle = 1 $.
Lattices form $ L=8 $ to $ L=16 $ were considered.
(b) same as (a) but for $\pht/t = 0$.  The calculations in this figure 
were carried out at $\Phi = 0$ (see Eq. \ref{Bound}).

\item (a) $ S(L/2,L/2)$ versus $1/L$ for several values of $\pht/t$. The
solid lines correspond to least square fits of the QMC data to the form
$1/L$.  Inset: $ S(L/2,L/2)$ versus $1/L$  at $\pht/t = 0.6$. The solid
is a least square fit to the form $L^{-\alpha}$.
(b) Staggered moment  as obtained from (a) versus $\pht/t$. 
The data point at $\pht/t = 0$ is taken from reference \cite{White}.
At $\pht/t = 0.3$, we were unable to distinguish $m$ from zero within our
statistical uncertainty.   The solid line is a guide to the eye. 
The calculations in this figure
were carried out at $\Phi = 0$ (see Eq. \ref{Bound}).

\end{list}


\begin{thebibliography}{99}

\bibitem{Scalapino} D.J. Scalapino {\it Does the Hubbard model have the right
stuff?} Proceedings of the International School of Physics, Enrico Fermi,
Course CXXI,   edited by R.A. Broglia and J.R. Schrieffer, North-
Holland, (1994).   D.J. Scalapino, preprint, cond-mat/9606218,
cond-mat/9606217 and references therein. 

\bibitem{Su} W.P. Su, J.R. Schrieffer, and A.J. Heeger, Phys. Rev. B
{\bf 22}, 2099 (1980).

\bibitem{Hirsch1} J.E. Hirsch, Phys. Rev. B{\bf 35}, 8726, (1987).

\bibitem{Imada1} M. Imada, Prog. Theor. Phys. Suppl. {\bf 113}, 203, (1993).

\bibitem{Hirsch} J.E. Hirsch and E. Fradkin, Phys. Rev B {\bf 27}, 4302,
(1983). 

\bibitem{Koonin} G. Sugiyama and S.E. Koonin, Anals of Phys.{\bf 168}
(1986) 1.
 
\bibitem{Sandro} S. Sorella, S. Baroni, R. Car, And M. Parrinello,
Europhys. Lett. {\bf 8} (1989) 663.
S. Sorella, E. Tosatti, S. Baroni, R.
Car, and M. Parinello, Int. J. Mod. Phys. B{\bf 1} (1989) 993.

\bibitem{Assaad2} F.F. Assaad, M. Imada and D.J. Scalapino, unpublished.

\bibitem{Kohn} W. Kohn, Phys. Rev. {\bf 133A}, 171, (1964).

\bibitem{Byers} N. Byers and C.N. Yang, Phys Rev. Lett.{\bf 7}, 46, (1961).

\bibitem{Yang}  C.N. Yang, Reviews of Mod. Phys. {\bf 34}, 694 (1962).

\bibitem{Assaad1}  F.F. Assaad, W. Hanke and D.J. Scalapino, Phys. Rev.
Lett. {\bf 71}, 1915 (1993),  Phys. Rev. B {\bf 49}, 4327 (1994).

\bibitem{Nejat} N. Bulut and D.J. Scalapino, Phys. Rev. B {\bf 45} 2371,
(1992).

\bibitem{Imada} M. Imada,  J. Phys. Soc. of Jpn. {\bf 64}, 2954 (1995).

\bibitem{Furukawa}  N. Furukawa and M. Imada, J. Phys. Soc. Jpn. {\bf 62},
2557, (1993). N. Furukawa, F.F. Assaad and  M. Imada, J. Phys. Soc. Jpn
{\bf 65}, 2339, (1996).

\bibitem{Assaad}  F.F. Assaad and M. Imada, Phys. Rev. Lett. {\bf 76}, 
3176, (1996).

\bibitem{White} S.R. White et al.  Phys. Rev. B{\bf 40},
506, (1989).



\end{thebibliography}
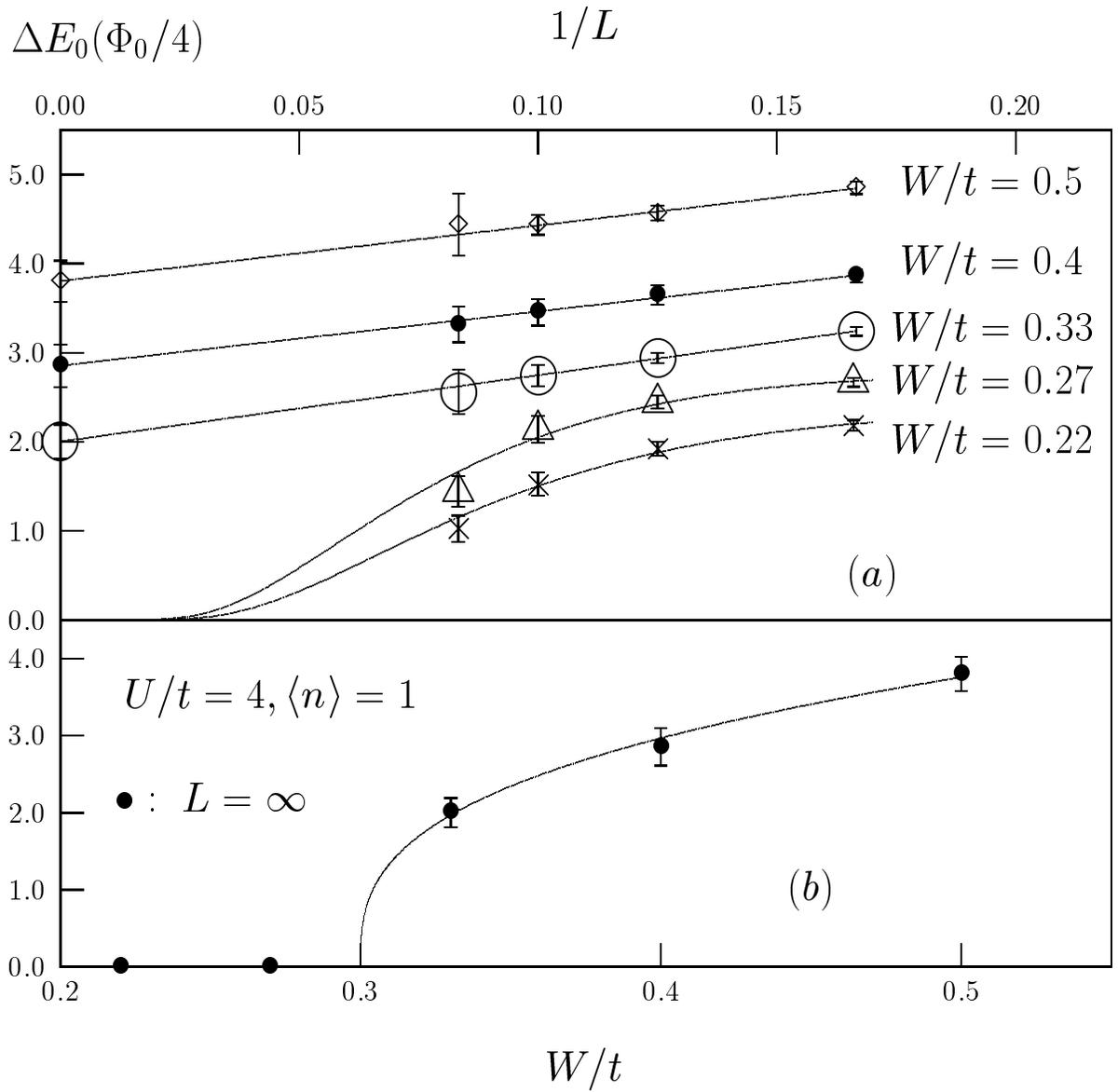
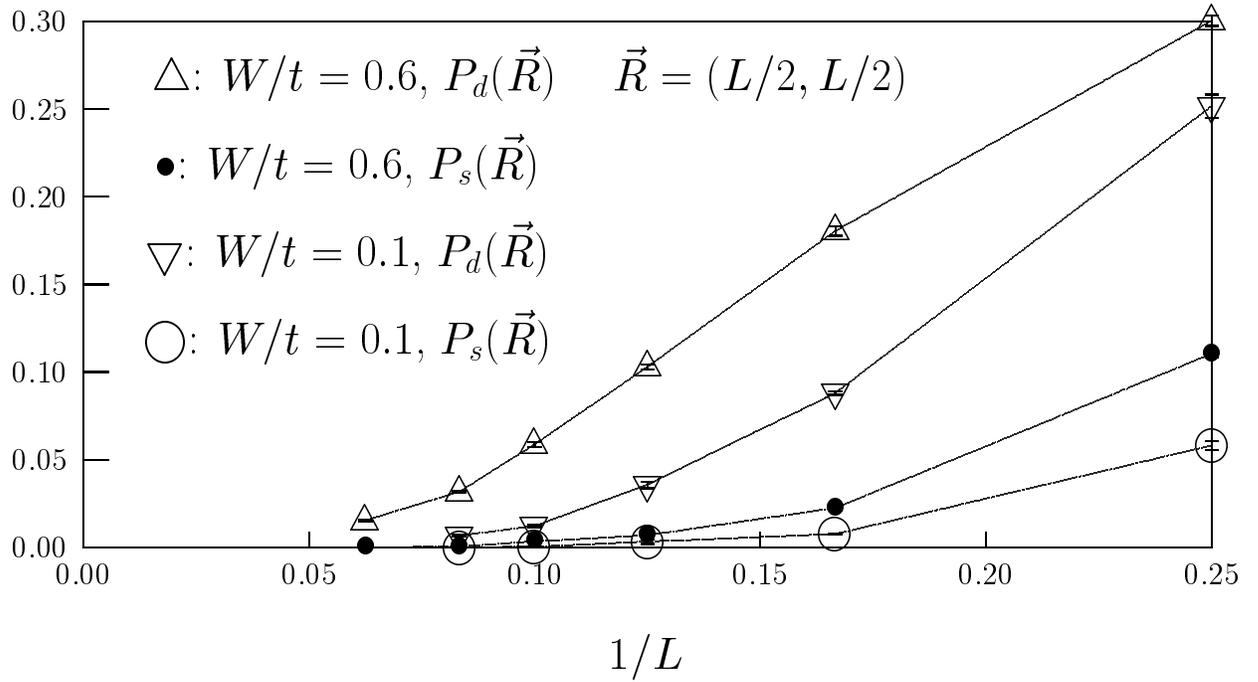
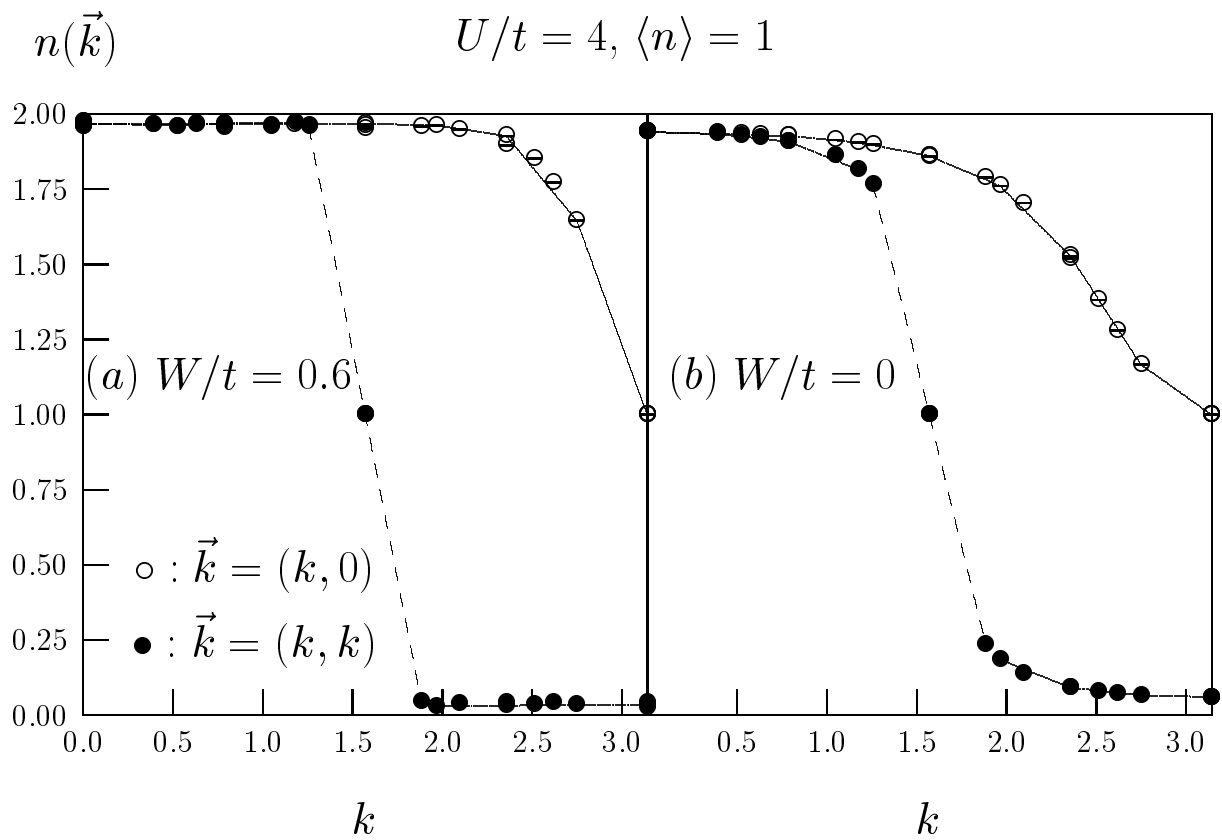
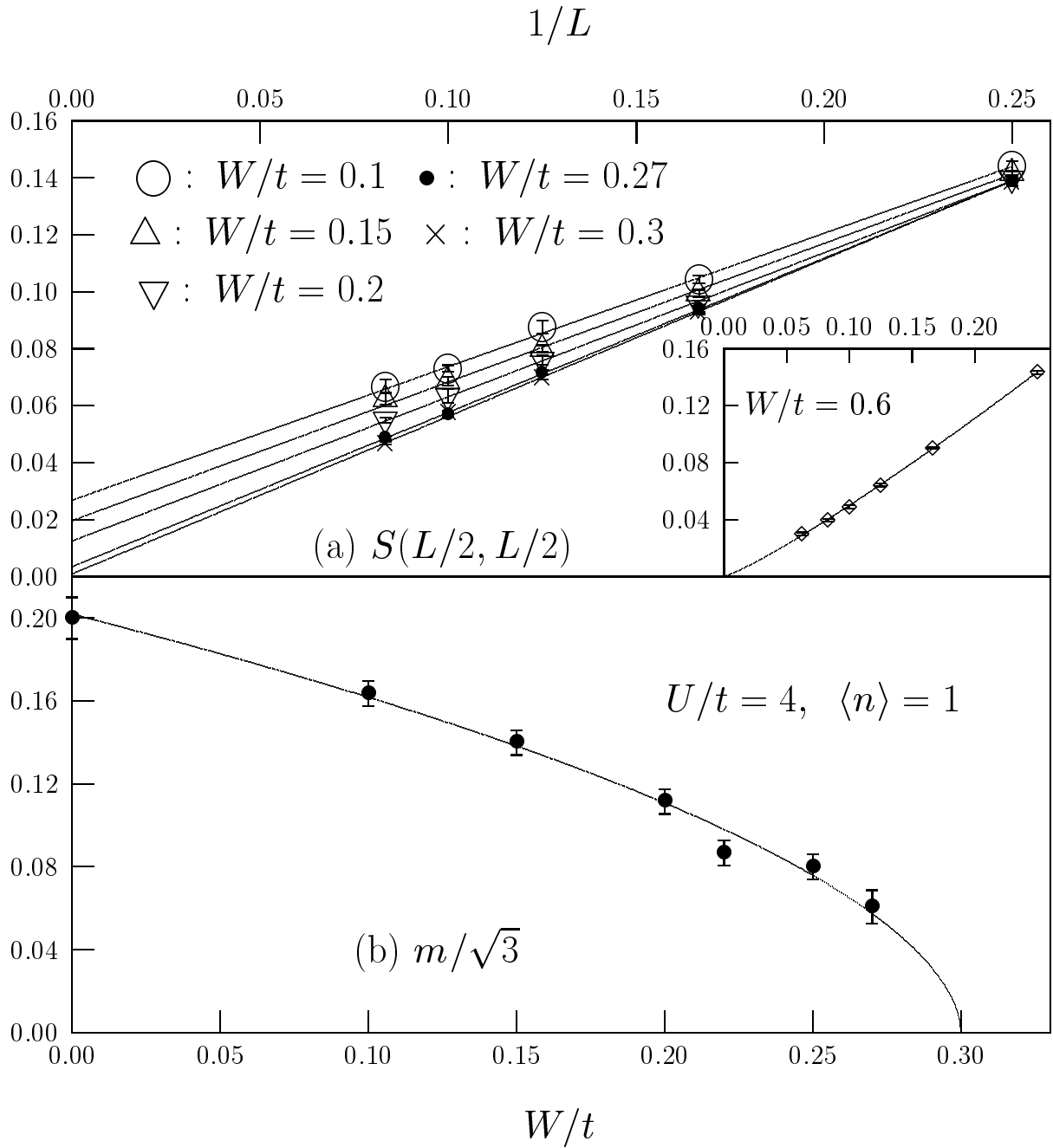
\end{document}